# A simple derivation of $E = mc^2$

Peter M. Brown
E-mail: pmb61@hotmail.com

Abstract – Einstein's 1905 derivation of $E = mc^2$ has been criticized for being circular. Although such criticism have been challenged it is certainly true that the reasoning in Einstein's original derivation is not at all obvious. Einstein's original derivation could be been made clearer. This article describes a clear way of doing so.

## I - Introduction

In Einstein's 1905 paper *On the Electrodynamics of Moving Bodies* [1] he derived an expression for the relationship for the energy of light wave as observed from two different inertial frames of reference. In another paper in the same issue the journal Einstein's paper *Does the inertia of a body depend on its energy content?* appeared [2]. In it he used the aforementioned energy relationship to derive what has become written as $E = mc^2$ which relating to the change in the mass of the body due to a decrease in. His arguments were later criticized, notably by H. Ives, for being circular [3]. However even though such criticisms have been challenged they do demonstrate the difficulty in following Einstein's logic. In this paper I provide a simple derivation in the spirit of that given by Einstein but which is much easier to understand. Einstein's derivation was based on the change in kinetic energy which results from a decrease in the bodies mass where he mass is the $m$ in $K = mv^2/2$. This relation provides a secondary meaning for the body's mass. By this I mean that the $m$, the inertial mass, is defined as the $m$ in $p = mv$. The relation $K = mv^2/2$ is a derived relation and has a meaning not directly associated with the body's inertial mass. Although this is not an objectionable usage of the inertial mass, a more direct usage would be more satisfactory. Especially if the proof becomes easier to follow. I present below what I believe to be a truly intuitively obvious derivation.

## II - Derivation of $E = mc^2$ based on conservation of momentum

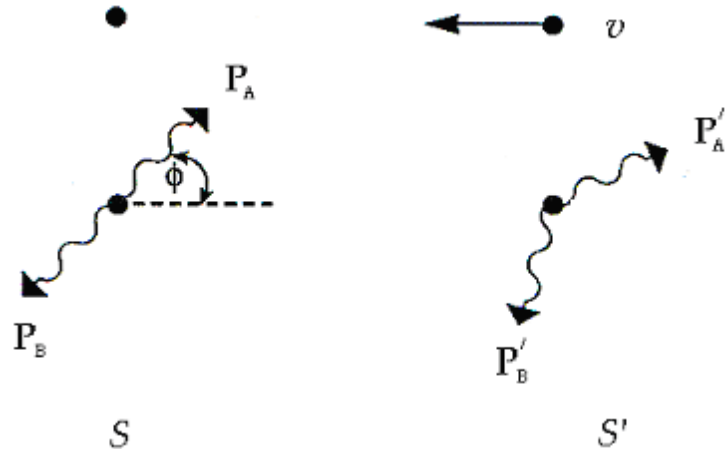

Consider a body at rest in an inertial frame, *S*, as shown above. The body emits two photons, *a* and *b*, of equal energy in opposite directions. The total energy of the two photons, as measured in *S*, is *E*. Due to the conservation of momentum, the emitting body must remain at rest in *S*. It follows that the velocity must be constant in *S′*. As viewed from *S′* before the photons are emitted there is only a single body, the total momentum being that of the body which is defined as

(1)   $\mathbf{P'}_i = -m'_i\, v\, \mathbf{e}_x$

where $m'_i$ is the inertial mass of the body before the emission. After the body emits the two photons the total momentum will be still be conserved and will have the value

(2)   $-m'_i\, v\, \mathbf{e}_x = -m'_f\, v\, \mathbf{e}_x + \mathbf{P'}_a + \mathbf{P'}_b = -m'_f\, v\, \mathbf{e}_x + \left(P'_{ax} + P'_{bx}\right)\mathbf{e}_x + \left(P'_{ay} + P'_{by}\right)\mathbf{e}_y$

where $m'_f$ is the inertial mass of the body after the emission, $\mathbf{P'}_a$ and $\mathbf{P'}_a$ are the momenta of photon's *a* and *b*, respectively, as measured in *S′*. Equating the x-components of Eq. (2) we find

(3)   $-\Delta m' v\, \mathbf{e}_x = -\left(m'_i - m'_f\right)\mathbf{e}_x = P'_{ax} + P'_{bx}$

Einstein showed that the energies, as measured in *S′*, of photons *a* and *b* are [1]

(4a)   $E'_a = \dfrac{\gamma E}{2c}\left(1 - \beta \cos\varphi\right)$

(4b)   $E'_b = \dfrac{\gamma E}{2c}\left(1 + \beta \cos\varphi\right)$

(5)   $E'_a + E'_b = \gamma E = E'$

The x-components of the momenta in Eq. (3) are related to the energies in Eq. (4) according to the relation $E = Pc$ which is deduced from Maxwell's equations. The x-components of the photon momenta, in $S'$, are

(6a) $\quad P'_{ax} = \dfrac{E'_a}{c} \cos\varphi'_a$

(6b) $\quad P'_{bx} = \dfrac{E'_b}{c} \cos\varphi'_b$

The cosines are found be transforming the components of the velocity of light from $S$ to $S'$ using the velocity transformation relations. Referring to the diagram below

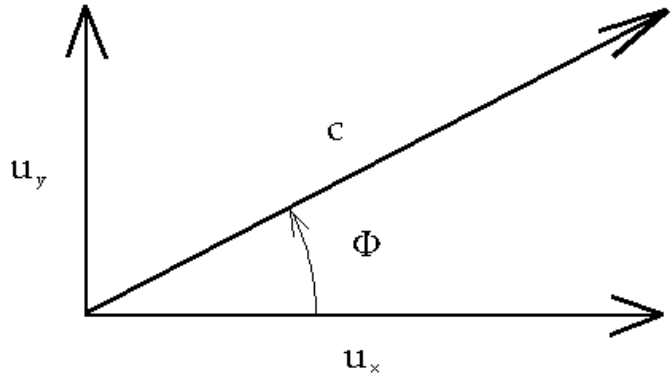

(7a) $\quad \cos\varphi_a = \dfrac{u_{ax}}{c}$

(7a) $\quad \cos\varphi_b = -\dfrac{u_{bx}}{c}$

where $\cos\varphi_b = -\cos\varphi_a = -\cos\varphi$. The velocity transformation relation for the x component of velocity is [1]

(8) $\quad u'_x = \dfrac{u_x - v}{1 - u_x v/c^2}$

Using Eq. (6) and (7) gives

(9a) $\quad \cos\varphi_a' = \dfrac{u'_{ax}}{c} = \dfrac{\cos\varphi_a - a}{1 - a\cos\varphi_a} = \dfrac{u_{ax}/c - v}{1 - (u_{ax}/c)a} = \dfrac{\cos\varphi - a}{1 - a\cos\varphi}$

(9b) $\quad \cos\varphi_b' = \dfrac{u'_{bx}}{c} = \dfrac{\cos\varphi_b - a}{1 - a\cos\varphi_b} = -\dfrac{\cos\varphi + a}{1 + a\cos\varphi}$

We are now ready to evaluate the x-component of the photon momenta

(10a)　$P'_{ax} = \left[ \dfrac{\gamma E}{2c}\left(1 - \beta\cos\varphi\right)\right]\left[\dfrac{\cos\varphi - \beta}{1 - \beta\cos\varphi}\right] = \dfrac{\gamma E}{2c}\left(\cos\varphi - \beta\right)$

(10b)　$P'_{bx} = \left[ \dfrac{\gamma E}{2c}\left(1 + \beta\cos\varphi\right)\right]\left[-\dfrac{\cos\varphi + \beta}{1 + \beta\cos\varphi}\right] = -\dfrac{\gamma E}{2c}\left(\cos\varphi + \beta\right)$

(11)　$-\overset{..}{A}\overset{..}{A}m' = P'_{ax} + P'_{bx} = \dfrac{aE}{2c}\left(\cos\varphi - \hat{a}\right) - \dfrac{aE}{2c}\left(\cos\varphi + \hat{a}\right) = -\dfrac{aEv}{c^2}$

Upon substituting Eq. (5) we get

(12)　$-\Delta m'v = P'_{ax} + P'_{bx} = \dfrac{\gamma E}{2c}\left(\cos\varphi - \beta\right) - \dfrac{\gamma E}{2c}\left(\cos\varphi + \beta\right) = -\dfrac{E'v}{c^2}$

We thus arrive at our final result

(13)　$E' = \Delta m' c^2$

Therefore the inertial mass changes with a change in energy of the body. Since this relation holds for all $v < c$ it therefore holds in the limiting case of $v = 0$. We can therefore identify the energy as the energy of the radiation as emitted in the frame in which the body is at rest and interpret $\Delta m$ as a change in rest mass.

## III – Conclusion

The derivation I've given here seems like the most obvious approach to the equivalence of mass and energy. The spirit of the derivation is nearly identical to Einstein's but is more direct. Since mass is defined as being the ratio of momentum and velocity it seems quite natural to use the principle of momentum conservation to derive the relation. Very little is needed beyond the expression for the energy of the light in the moving and the velocity transformation relation, both of which are available in Einstein's 1905 paper. The derivation is also easy to follow and is readily accessible to those who wish to understand Einstein's original derivation since it serves as step in that direction since the spirit of the derivation is the same as Einstein's original derivation.